\documentclass[twocolumn]{IEEEtran}

\usepackage{amsmath,epsfig,amssymb,verbatim,amsopn,cite,multirow}
\usepackage{amsthm}
\usepackage{balance}
\usepackage{multirow}
\usepackage[usenames,dvipsnames]{color}
\usepackage[all]{xy}  
\usepackage{url}
\usepackage{amsfonts}
\usepackage{amssymb}
\usepackage{epsfig}
\usepackage{epstopdf}
\usepackage{bm}
\usepackage{balance}
\usepackage{graphicx}
\usepackage{footnote}
\usepackage{algorithm,algorithmic}
\usepackage{makecell}

\usepackage[margin=15.4mm,top=17.1mm,bottom=25.5mm]{geometry}

\newtheorem{Theorem}{Theorem}

  {\proof}{\proofend}

\newcommand{\qa}{{\bf a}}

\newcommand{\qe}{{\bf e}}

\newcommand{\qg}{{\bf g}}

\newcommand{\qw}{{\bf w}}
\newcommand{\qx}{{\bf x}}

\newcommand{\qB}{{\bf B}}

\newcommand{\qG}{{\bf G}}

\newcommand{\qI}{{\bf I}}

\newcommand{\ettall}{\emph{et al.}}

\newcommand{\dl}{\mathtt{dl}}

\newcommand{\UE}{\mathtt{I}}
\newcommand{\sn}{\mathtt{E}}

\DeclareMathOperator{\ETAI}{\boldsymbol{\eta}^{\mathtt{I}}}
\DeclareMathOperator{\ETAE}{\boldsymbol{\eta}^{\mathtt{E}}}
\DeclareMathOperator{\aaa}{\mathbf{a}}

\DeclareMathOperator{\MM}{\mathcal{M}}
\newcommand{\Sens}{\mathcal{L}}
\DeclareMathOperator{\K}{\mathcal{K}}

\DeclareMathOperator{\C}{\mathbb{C}}

\newcommand{\Ntx}{N}
\newcommand{\Nrx}{N}

\newcommand{\PZF}{\mathsf{PZF}}
\newcommand{\PMRT}{\mathsf{PMRT}}

\newcommand{\Sn}{\sigma_n^2}

\newcommand{\wimk}{\qw_{\mathrm{I},mk}}
\newcommand{\weml}{\qw_{\mathrm{E},m\ell}}

\newcommand{\wimkp}{\qw_{\mathrm{I},mk'}}

\newcommand{\wemlp}{\qw_{\mathrm{E},m\ell'}}
\newcommand{\Ghms}{\hat{\qG}_m^{\sn}}
\newcommand{\Ghmu}{\hat{\qG}_m^{\UE}}
\newcommand{\Snn}{\sigma_n^2}
\newcommand{\Ex}{\mathbb{E}}

\newcommand{\yik}{y_{\mathtt{I},k}}

\newcommand{\bsHz}{\text{[bit/s/Hz]}}

\newcommand{\HEQoS}{f(\omega_{\ell})}
\newcommand{\SEQoS}{\mathcal{S}_\dl^o}
\newcommand{\tilSEQoS}{\tilde{\mathcal{S}}_\dl^o}
\newcommand{\barSEQoS}{\bar{\mathcal{S}}_\dl^o}

\newcommand{\tgmkue}{\tilde{\qg}_{mk}^{\UE}}
\newcommand{\tgmls}{\tilde{\qg}_{m\ell}^{\sn}}

\newcommand{\gmkue}{\qg_{mk}^{\UE}}

\newcommand{\hgmkue}{\hat{\qg}_{mk}^{\UE}}

\newcommand{\hgmls}{\hat{\qg}_{m\ell}^{\sn}}

\newcommand{\gmls}{\qg_{m\ell}^{\sn}}

\newcommand{\gtilmls}{\tilde{\qg}_{m\ell}^{\sn}}

\newcommand{\ghmons}{\hat{\qg}_{m1}^{\sn}}
\newcommand{\ghmLs}{\hat{\qg}_{mL}^{\sn}}
\newcommand{\ghmonue}{\hat{\qg}_{m1}^{\UE}}
\newcommand{\ghmKdue}{\hat{\qg}_{mK_d}^{\UE}}
\newcommand{\ghmkue}{\hat{\qg}_{mk}^{\UE}}
\newcommand{\gtilmkue}{\tilde{\qg}_{mk}^{\UE}}

\newcommand{\gamuemk}{\gamma_{mk}^{\UE}}

\newcommand{\gamsml}{\gamma_{m\ell}^{\sn}}

\newcommand{\betamkue}{\beta_{mk}^{\UE}}

\newcommand{\betamls}{\beta_{m\ell}^{\sn}}

\newcommand{\etamkI}{\eta_{mk}^{\mathtt{I}}}

\newcommand{\etamkIn}{\eta_{mk}^{{\mathtt{I}}^{(n)}}}
\newcommand{\etamkpI}{\eta_{mk'}^{\mathtt{I}}}
\newcommand{\etamlE}{\eta_{m\ell}^{\mathtt{E}}}

\newcommand{\etamlEn}{\eta_{m\ell}^{{\mathtt{E}}^{(n)}}}
\newcommand{\etamlpE}{\eta_{m\ell'}^{\mathtt{E}}}

\newcommand{\etamIn}{\eta_{m}^{\mathtt{I}(n)}}
\newcommand{\etamEn}{\eta_{m}^{\mathtt{E}(n)}}

\newcommand{\umln}{u_{m\ell}^{(n)}}
\newcommand{\uml}{u_{m\ell}}

\newcommand{\Sto}{{\mathtt{S}_2}}
\newcommand{\Son}{{\mathtt{S}_1}}
\newcommand{\SEk}{\mathrm{SE}_k}
\newcommand{\SINRk}{\mathrm{SINR}_k}

\newcommand{\xik}{x_{\mathtt{I},k}}
\newcommand{\xikp}{x_{\mathtt{I},k'}}
\newcommand{\xel}{x_{\mathtt{E},\ell}}

\DeclareMathOperator{\BOmega}{\boldsymbol{\omega}}

\title{ Cell-free Massive MIMO and SWIPT: Access Point Operation Mode Selection and Power Control}

\author{Mohammadali Mohammadi$^\dag$, Le-Nam Tran$^\ddag$, Zahra Mobini$^\dag$, Hien Quoc Ngo$^\dag$, and  Michail Matthaiou$^\dag$\\
\small{
$^\dag$Centre for Wireless Innovation (CWI), Queen's University Belfast, U.K.\\
$^\ddag$School of Electrical and Electronic Engineering,
University College Dublin, Dublin 4, Ireland\\
Email:\{m.mohammadi, zahra.mobini, hien.ngo, m.matthaiou\}@qub.ac.uk, nam.tran@ucd.ie
}}\normalsize
\allowdisplaybreaks

\begin{document}

\bstctlcite{IEEEexample:BSTcontrol}
\maketitle
\begin{abstract} 
This paper studies cell-free massive multiple-input multiple-output (CF-mMIMO) systems incorporating  simultaneous wireless information and power transfer (SWIPT) for separate information users (IUs) and energy users (EUs) in Internet of Things (IoT) networks. To optimize both the spectral efficiency (SE) of IUs and harvested energy (HE) of EUs, we propose a joint access point (AP) operation mode selection and power control design, wherein certain APs are designated  for energy transmission to  EUs, while others  are dedicated to information transmission to  IUs. We investigate the problem of maximizing the total HE   for EUs, considering constraints on SE for individual IUs and minimum HE for individual EUs. Our numerical results showcase that the proposed AP operation mode selection algorithm can provide up to $76\%$ and $130\%$ performance gains over  random AP operation mode selection with and without power control, respectively.  

\let\thefootnote\relax\footnotetext{ This project has received funding from the European Research Council (ERC) under the European Union's Horizon 2020 research and innovation programme (grant agreement No. 101001331). The work of H. Q. Ngo was supported by the U.K. Research and Innovation Future Leaders Fellowships under Grant MR/X010635/1. The work of L.-N. Tran was supported by a Grant from Science Foundation Ireland under Grant number 17/CDA/4786.}
\end{abstract}


\vspace{-2em}
\section{Introduction}
Wireless power transfer (WPT) is an innovative technology which has experienced tremendous advancements over the past decade, enabling disruptive applications, such as  battery-less sensors, passive wireless sensors, and  IoT devices. By harvesting energy from the radio-frequency (RF) signals, broadcasted by  ambient/dedicated wireless transmitters, WPT can support the operation of energy-constrained wireless devices~\cite{Clerckx:JSAC:2019}. Nevertheless, the main challenge for WPT is the fast decline in energy transfer efficiency over  distance due to the severe path loss. To address this problem, researchers have studied MIMO systems, specially massive MIMO, along with energy beamforming techniques for their ability to focus highly directional RF signal power towards  user equipments (UEs)~\cite{Yang:JSAC:2015}. Despite all the progress achieved up to date,  energy harvesting capabilities for cell-boundary UEs remain intrinsically limited, potentially leading to a critical fairness concern among UEs.  

The aforementioned challenges can be effectively addressed in CF-mMIMO, where the APs are spatially distributed throughout the coverage area. This reduces the distance between UEs and nearby APs, resulting in greater macro-diversity and lower path loss~\cite{Hien:cellfree}, thereby making WPT more feasible. Consequently, various research efforts have been endeavored to investigate the  WPT performance in CF-mMIMO networks~\cite{Shrestha:GC:2018,Wang:JIOT:2020,Demir:TWC:2021,Femenias:TCOM:2021,Zhang:IoT:2022}. Shrestha \ettall~\cite{Shrestha:GC:2018} investigated the performance of SWIPT  in CF-mMIMO, where  IUs and EUs are located separately.  Wang~\ettall~\cite{Wang:JIOT:2020} considered minimizing the total transmit power for wirelessly-powered cell-free IoT with a linear energy harvesting model. Demir~\ettall~\cite{Demir:TWC:2021}  studied the power control and large-scale fading decoding weights design for maximizing the minimum UL SE for DL WPT-assisted CF-mMIMO. Femenias \ettall~\cite{Femenias:TCOM:2021} considered a  CF-mMIMO with separated EUs and IUs, developing a coupled UL/DL power control algorithm to optimize the weighted signal-to-interference-plus-noise ratio (SINR) of EUs. Zhang \ettall~\cite{Zhang:IoT:2022} proposed a max–min power control policy aiming to achieve uniform harvested energy and DL SE across all sensors in a CF-mMIMOSWIPT IoT network. 

All  works discussed above~\cite{Shrestha:GC:2018,Wang:JIOT:2020,Demir:TWC:2021,Femenias:TCOM:2021,Zhang:IoT:2022} have demonstrated that CF-mMIMO, thanks to its user-centric architecture,  can provide seamless energy harvesting opportunity for all EUs. However, even  with an optimal power control design, all these designs would still suffer from the  fundamental limitation in simultaneously increasing both the SE and HE for separate EUs and IUs. This is due to the inefficient use of available resources, as DL WPT towards EUs and  DL (UL) wireless information transfer (WIT) towards IUs (APs) occur over orthogonal time slots. A straightforward approach to enhance both the SE and HE would be to  deploy a large number of APs, but this is not energy efficient due to the large fronthaul burden and transmit power requirements~\cite{Hien:TGCN:2018}. To overcome this issue, we propose a novel network architecture that jointly designs the AP operation mode selection and power control strategy to maximize the HE under the constraints on per-IU SE and per-EU HE.  Specifically, relying on the long-term channel state information (CSI), the APs are divided into information transmission APs (termed as I-APs) and energy transmission APs (termed as E-APs), which  simultaneously serve IUs and EUs over the whole time slot period. While this new architecture provides EUs with an opportunity to harvest energy from all APs, it also creates increased interference at the IUs due to concurrent E-AP transmissions. To deal with this problem, we hereafter apply local partial zero-forcing (PZF) precoding and protective maximum ratio transmission (PMRT) to the I-APs and E-APs, respectively, to guarantee full protection for the IUs against  energy signals intended for EUs. The main contributions of this paper are:
\begin{itemize}
    \item We derive  closed-form expressions for the DL SE and HE of the IUs and EUs, respectively. Then, we formulate the problem of joint AP operation mode selection and power control, considering  per-AP power constraints as well as  SE and HE constraints for  IUs and EUs, respectively.   
    \item We develop an iterative algorithm based on successive convex approximation (SCA), to solve the complicated binary non-convex optimization problem.  
    \item Our numerical results demonstrate that the proposed architecture  improves significantly the energy harvesting performance compared to the benchmark schemes. For specific SE and HE requirements, it boosts the energy harvesting efficiency by an order of magnitude, compared to  conventional designs via orthogonal transmission through time division between information and energy transfer. 
\end{itemize}     

\textit{Notation:} We use bold upper case letters to denote matrices, and lower case letters to denote vectors. The superscript $(\cdot)^H$ stands for the conjugate-transpose.  
A zero mean circular symmetric complex Gaussian variable having variance $\sigma^2$ is denoted by $\mathcal{CN}(0,\sigma^2)$. Finally, $\mathbb{E}\{\cdot\}$ denotes the statistical expectation.

\vspace{-0.5em}
\section{System model}~\label{sec:Sysmodel}
\begin{figure}[t]
	\centering
	\vspace{0em}
	\includegraphics[width=60mm]{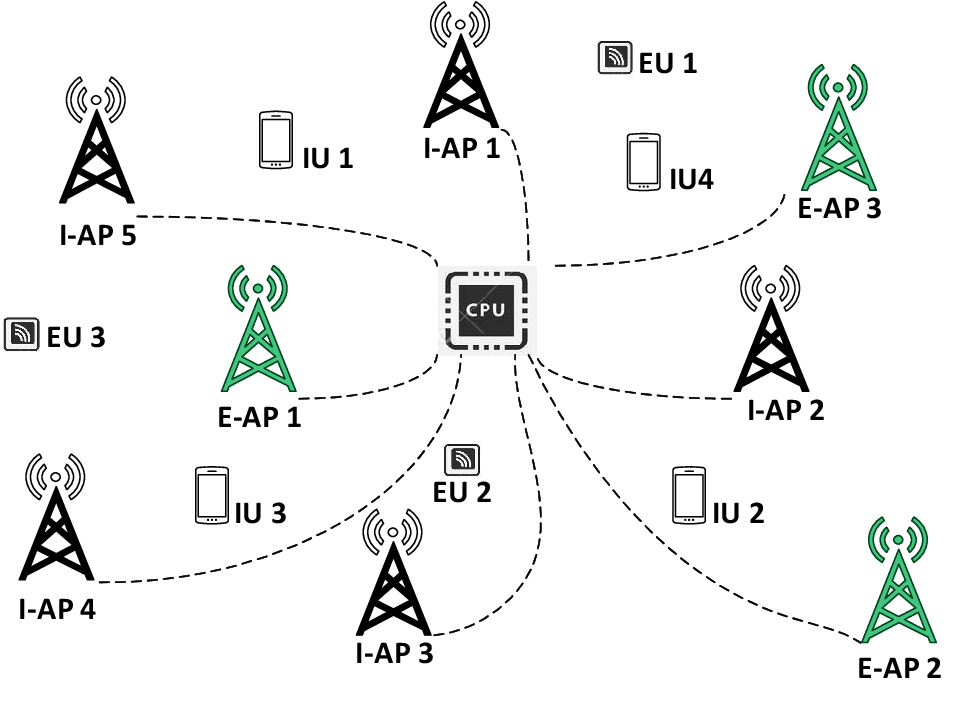}
	\vspace{-1.1em}
	\caption{Illustration of the proposed SWIPT CF-mMIMO system.}
	\vspace{0.2em}
	\label{fig:systemmodel}
\end{figure}

We consider a CF-mMIMO system  under time division duplex operation, where $M$ APs serve   $K_d$ IUs and $L$ EUs with energy harvesting capabilities in the DL. 
Each IU and EU are equipped with one single antenna, while each AP is equipped with $N$ antennas. All APs, IUs, and EUs operate as half-duplex devices. For notational simplicity, we define the sets $\mathcal{M}\triangleq\{1,\ldots,M\}$, $\K_d\triangleq \{1,\dots,K_d\}$ and  $\mathcal{L}\triangleq\{1,\ldots,L\}$ as collections of indices of the APs, IUs, and EUs, respectively. As shown in Fig.~\ref{fig:systemmodel},  information and energy transmissions take place simultaneously and within the same frequency band. The AP operation mode selection approach is designed according to the network requirements, determining whether an AP is dedicated to information or energy transmission. 
The IUs receive information from a group of the APs (I-APs), while the EUs harvest energy from the remaining APs (E-APs).
The EUs utilize  the harvested energy to transmit pilots and data.  Each coherence block includes two phases: 1) UL training for channel estimation; 2) DL WIT and WPT. We assume a quasi-static channel model, with each channel coherence interval spanning a duration of $\tau_c$ symbols. The duration of the training is denoted as $\tau$, while the duration of DL WIT and WPT is $(\tau_c-\tau)$.

\vspace{-1em}
\subsection{Uplink Training for Channel Estimation}\label{phase:ULforCE}
The channel vector between the $k$-th IU ($\ell$-th EU) and the $m$-th AP is denoted by $\gmkue\in\C^{\Ntx \times 1}$ ($\gmls\in\C^{\Nrx \times 1}$), $\forall k \in \K_d$, ($\ell \in \Sens$) and $\forall m \in \MM$. It is modeled as $ \gmkue=\sqrt{\betamkue}\tgmkue,~(\gmls=\sqrt{\betamls}\tgmls) $, where $\betamkue$ ($\betamls$) is the large-scale fading coefficient and $\tgmkue\in\C^{\Ntx \times 1}$ ($\tgmls\in\C^{\Ntx \times 1}$) is the small-scale fading vector, whose elements are independent and identically distributed $\mathcal{CN} (0, 1)$ random variables. 

In each coherence block of length $\tau_c$, all IUs and EUs are assumed to transmit their pairwisely orthogonal pilot sequences of length $\tau$ to all the APs, which requires $\tau\geq K_d + L$. At AP $m$, $\gmkue$  and $\gmls$ are estimated by using the received pilot signals and the minimum mean-square error (MMSE) estimation technique. By following~\cite{Hien:cellfree}, the MMSE estimates $\hgmkue$ and $\hgmls$ of $\gmkue$  and $\gmls$ are $\hgmkue \sim \mathcal{CN}(\boldsymbol{0},\gamuemk \mathbf{I}_N)$, and $\hgmls \sim \mathcal{CN}(\boldsymbol{0},\gamsml \mathbf{I}_N)$, respectively,  where $\gamuemk \triangleq \frac{{\tau\rho_t}(\betamkue)^2}{\tau\rho_t \betamkue+1}$, and $\gamsml \triangleq\frac{{\tau\rho_t}(\betamls)^2}{\tau\rho_t \betamls +1}$, while $\rho_t$ is the normalized signal-to-noise ratio (SNR) of each pilot symbol. Furthermore, the corresponding channel estimation errors are denoted by $\gtilmkue\sim\mathcal{CN}(\boldsymbol{0},(\betamkue-\gamuemk)\qI_N)$ and $\gtilmls\sim\mathcal{CN}(\boldsymbol{0},(\betamls-\gamsml)\qI_N)$.

\vspace{-0.7em}
\subsection{Downlink Information and Power Transmission}
In this phase, the APs are able to switch between the information and energy transmission functionalities and use different precoders among the PZF and PMRT to serve IUs and EUs, respectively.  The decision of which operation mode is assigned to each AP is optimized to achieve the highest average sum-HE at the EUs considering the SE requirements of the IUs, as will be discussed in Section~\ref{sec:optimization}. Note that the AP operation mode selection is performed on a large-scale fading timescale which changes slowly with time. The binary variables to indicate the operation mode selection for each AP $m$ are defined as
\vspace{-0.7em}
\begin{align}
\label{a}
a_{m} &\triangleq
\begin{cases}
  1, & \text{if AP $m$ operates as I-AP}\\
  0, & \text{if AP $m$ operates as E-AP.}
\end{cases}
\end{align}
In DL, all I-APs  aim to transmit a data symbol $\xik$, with $\Ex\big\{\vert\xik\vert^2\big\}=1$ to IU $k\in\K_d$. At the same time, all E-APs  transmit energy symbol $\xel$, with $\Ex\big\{\vert\xel\vert^2\big\}=1$ to IU $\ell\in\Sens$.  The vector of the transmitted signal from AP $m$ is  $\qx_{m} = \sqrt{a_m} \qx_{\mathtt{I},m} +\sqrt{(1-a_m)}\qx_ {\mathtt{E},m}$,
where $\qx_{\mathtt{I},m}=\sqrt{\rho}\sum_{k=1}^{K_d}\sqrt{\etamkI}\wimk \xik$ and $\qx_ {\mathtt{E},m}=\sqrt{\rho}\sum_{\ell=1}^{L}  \sqrt{\etamlE}\weml \xel$, with $\rho$ being the maximum normalized DL SNR; $\wimk \in \C^{N\times 1}$ and $\weml\in \C^{N\times 1}$ are the precoding vectors for IU $k$ and EU $\ell$, respectively, with $\Ex\big\{\big\Vert\wimk\big\Vert^2\big\}=1$ and $\Ex\big\{\big\Vert\weml\big\Vert^2\big\}=1$. Note that AP $m$ can only transmit either $\qx_{\mathtt{I},m}$ or $\qx_ {\mathtt{E},m}$, depending on its assigned operation mode. Moreover, $\etamkI$ and $\etamlE$ are the DL
power control coefficients chosen to satisfy the power constraint at each AP, given by
\vspace{-0.0em}
\begin{equation}
a_m\Ex\big\{\big\Vert \qx_{\mathtt{I},m}\big\Vert^2\big\}+(1-a_m)\Ex\big\{\big\Vert \qx_{\mathtt{E},m}\big\Vert^2\big\}\leq \rho.
\end{equation}

\vspace{-0.5em}
\subsection{Downlink Spectral Efficiency and Average Harvested Energy}
By invoking the bounding technique in~\cite{Hien:cellfree}, known as the hardening bound, we derive a lower bound on the DL SE of the IU $k$. To this end, we express the received signal at IU $k$ as
\begin{align}~\label{eq:yi:hardening}
    \yik &=  \mathrm{DS}_k  \xik +
    \mathrm{BU}_k \xik 
         +\sum_{k'\in\K_d\setminus k}
     \mathrm{IUI}_{kk'}
     \xikp\nonumber\\
    &\hspace{1em}
    + \sum_{\ell\in\Sens}
     \mathrm{EUI}_{k\ell}\xel + n_k,~\forall k\in\K_d,
\end{align}
where $\mathrm{DS}_k \!= \!\! \sum_{m=1}^{M}\!\!\sqrt{\rho a_m\etamkI} \Ex\big\{\big(\gmkue\big)^H\wimk \big\}$, $\mathrm{BU}_k  \!\!= \!\!  \sum_{m=1}^{M}\!\sqrt{\rho a_m\etamkI}\Big(\big(\gmkue\big)^H\wimk \!-\!  \Ex\Big\{\big(\gmkue\big)^H\wimk \Big\} \Big)$, $\mathrm{IUI}_{kk'} \!=\! \sum_{m=1}^{M}\sqrt{\rho a_m\etamkpI} \big(\gmkue\big)^H\wimkp$, $\mathrm{EUI}_{k\ell}\!=\!\sum_{m=1}^{M} \sqrt{\rho b_m\etamlE}\big(\gmkue\big)^H\weml$, represent the desired signal ($\mathrm{DS}_k$),  the beamforming gain uncertainty ($\mathrm{BU}_k$), the interference caused by the $k'$-th IU ($\mathrm{IUI}_{kk'}$), and the interference caused by the $\ell$-th EU ($\mathrm{EUI}_{k\ell}$), respectively. Accordingly, by invoking the bounding techniquein~\cite{Hien:cellfree}, an achievable DL SE for IU $k$ in $\bsHz$ can be written as
\begin{align}~\label{eq:SEk:Ex}
    \mathrm{SE}_k(\aaa, \ETAI, \ETAE)
      &=
      \Big(1\!- \!\frac{\tau}{\tau_c}\Big)
      \log_2
      \left(
       1\! + \mathtt{SINR}_k (\aaa, \ETAI, \ETAE)
     \right),
\end{align}
where $ \ETAI = [\eta_{m1}^{\mathtt{I}}, \ldots, \eta_{mK_d}^{\mathtt{I}}]$, $ \ETAE = [\eta_{m1}^{\mathtt{E}}, \ldots, \eta_{mL}^{\mathtt{E}}]$,  and $\mathtt{SINR}_k(\aaa,  \ETAI, \ETAE)=$ 
\begin{align}~\label{eq:SINE:general}
       \!\frac{
                 \big\vert  \mathrm{DS}_k  \big\vert^2
                 }
                 {  
                 \Ex\big\{ \big\vert  \mathrm{BU}_k  \big\vert^2\big\} +
                  \!\!
                 \sum_{k'\neq k}
                  \Ex\big\{ \big\vert \mathrm{IUI}_{kk'} \big\vert^2\big\}
                  \! + \!
                   \!
                  \sum_{\ell=1}^{L}
                 \Ex \big\{ \big\vert  \mathrm{EUI}_{k\ell} \big\vert^2\big\}
                   +  1}.
\end{align}

To characterize the HE precisely, a non-linear energy harvesting model with the sigmoidal function is used. Therefore, the total HE at EU $\ell$ is given by~\cite{Boshkovska:CLET:2015}
 \vspace{0.1em}
  \begin{align}~\label{eq:NLEH}
  \Phi\big(\aaa,  \ETAE,\ETAI\big) = \frac{\Psi_{\ell}\big(\mathrm{E}_{\ell}(\aaa,  \ETAE, \ETAI)\big) - \phi \Omega }{1-\Omega}, ~\forall \ell\in\Sens,
 \end{align}
where  $\phi$ is the maximum output DC power, $\Omega=\frac{1}{1 + \exp(\xi \chi)}$ is a constant to guarantee a zero input/output response, while $\Psi\big(\mathrm{E}_{\ell}(\aaa,  \ETAE, \ETAI)\big)$  is the traditional logistic function, 
 \vspace{0.1em}
  \begin{align}~\label{eq:PsiEl}
     \Psi_{\ell}\big(\mathrm{E}_{\ell}(\aaa,  \ETAE,  \ETAI)\big) &\!=\!\frac{\phi}{1 \!+ \!\exp\Big(-\xi\big(\mathrm{E}_{\ell}(\aaa, \ETAE,\ETAI)\!-\! \chi\big)\Big)},
 \end{align}
where $\xi$ and $ \chi$ are constant related parameters that depend on the circuit. Moreover, $\mathrm{E}_{\ell}(\aaa, \ETAE,\ETAI)$ denotes the received RF energy at EU $\ell$, $\forall \ell\in\mathcal{L}$. We denote the average of received power as $ Q_{\ell}(\aaa, \ETAE, \ETAI) \!=(\tau_c-\tau)\Snn\Ex\big\{\mathrm{E}_{\ell}(\aaa, \ETAE, \ETAI)\big\}$, and 
\vspace{0.1em}
\begin{align}~\label{eq:El_average}
     &\Ex\Big\{\!\mathrm{E}_{\ell}(\aaa, \ETAE, \ETAI)\!\Big\}
     \!=\!
      \! {\rho}\!\!\sum_{m=1}^{M}\!\sum_{\ell'=1}^{L}\!\!
   {(1\!-\!a_m)\etamlpE} \Ex\big\{\!\big\vert\big(\gmls\big)^{\!\!H}\wemlp\big\vert^2\!\big\} 
   \nonumber\\
   &\hspace{3em} +{\rho}
   \sum_{m=1}^{M}\!\sum_{k=1}^{K_d}
   {a_m\etamkI}\Ex\big\{\big\vert\big(\gmls\big)^{\!H}\wimk\big\vert^2\big\} + 1.
\end{align}

We derive closed-form expressions for the proposed precoding scheme in the following subsection.
\vspace{-1.5em}
\subsection{Protective Partial Zero-Forcing}
We propose to utilize the protective PZF scheme at the APs, where local PZF precoding is deployed at the I-APs and PMRT is used at the E-APs. The principle behind this design is that ZF precoders work very well and nearly optimally for information transmission due to their ability to suppress the interuser interference \cite{Hien:cellfree}. On the other hand, MRT is shown to be an optimal beamformer for power transfer that maximizes the HE in the case when $N$ is large \cite{almradi2016performance}. Nevertheless, the IUs experience non-coherent interference from the energy signals transmitted to the EUs. To reduce this interference, MRT can be forced to take place in the orthogonal complement of the IUs' channel space. This design is called PMRT beamforming. 

We define the matrix of channel estimates for the $m$-th AP as $\hat{\qG}_m = \big[\Ghmu, \Ghms\big] \in\C^{N\times(K_d+L)}$, where  $\Ghmu = \big[\ghmonue, \cdots, \ghmKdue\big]$ denotes the estimate of all channels between AP $m$ and all IUs,  $\Ghms= \big[\ghmons, \cdots, \ghmLs\big]$ is the estimate of all channels between AP $m$ and all EUs. Now, the PZF and PMRT precoder at the $m$-th AP towards IU $k$ and EU $\ell$, can be expressed as
\vspace{-0.3em}
\begin{subequations}
 \begin{align}
    \wimk^{\PZF} &=\alpha_{\PZF,mk}
    { \Ghmu \Big(\big(\Ghmu\big)^H \Ghmu\Big) ^{-1} \qe_k^I}
    ,~\label{eq:wipzf}\\
        \weml^{\PMRT} &= \alpha_{\PMRT,m\ell}  \qB_m\Ghms\qe_{\ell}^{E},~\label{eq:wemrt}
\end{align}   
\end{subequations}
where $\qe_{k}^{I}$ ($\qe_{\ell}^{E}$) is the $k$-th column of $\qI_{K_d}$ ($\ell$-th column of $\qI_{L}$); $\alpha_{\PZF,mk} = \sqrt{(N-K_d)\gamuemk}$ and $\alpha_{\PMRT,m\ell} = \frac{1}{\sqrt{\big(N-K_d\big)\gamsml}}$
denote the precoding normalization factors; $\qB_m$ denotes the projection matrix onto the orthogonal complement of $\Ghmu$, i.e.,
\vspace{-0.5em}
\begin{align}
  \qB_m  = \qI_{\Ntx}  - \Ghmu \Big( \big(\Ghmu\big)^H \Ghmu\Big)^{-1}  \big(\Ghmu\big)^H,
\end{align}
which implies $\big(\ghmkue\big)^H \qB_m =0$.

In the following theorems, we provide closed-form expressions for the SE and average HE under protective PZF (i.e. PZF at I-APs and PMRT precoding at E-APs). 
The proof of the theorems are omitted due to the space limitation.

\begin{Theorem}~\label{Theorem:SE:PPZF}
The ergodic SE for the $k$-th IU, achieved by PZF precoding at the I-APs and PMRT at the E-APs is given by~\eqref{eq:SEk:Ex}, where the effective SINR is given in closed-form by~\eqref{eq:SINE:PPZF} at the top of the next page.
\begin{figure*}
\begin{align}~\label{eq:SINE:PPZF}
    \SINRk(\aaa, \ETAI, \ETAE) =
    \!\frac{
                  \rho \big(N-K_d\big)\Big(\sum_{m=1}^{M}\sqrt{ a_m\etamkI \gamuemk}  \Big)^2
                 }
                 { \rho \sum_{m=1}^{M}
                 \sum_{k'=1}^{K_d}
  a_m\etamkpI 
  \Big(\betamkue\!-\!\gamuemk\Big)
                  \! + \!
                 \rho
                 \sum_{m=1}^{M}
                 \sum_{\ell=1}^{L}
   { (1\!-\!a_m)\etamlE} \Big(\betamkue\!-\!\gamuemk\Big)
                   \!+\!  1}.
\end{align}
	\hrulefill
	\vspace{-4mm}
\end{figure*}
\end{Theorem}


\begin{Theorem}~\label{Theorem:RF:PPZF}
The average HE for the $\ell$-th EU, achieved by the PMRT precoding at the E-APs and PZF at I-APs is given by~\eqref{eq:El_average:PPZF} at the top of the next page.
\begin{figure*}
\begin{align}~\label{eq:El_average:PPZF}
    Q_{\ell} (\aaa, \ETAI, \ETAE)
 &=
     (\tau_c-\tau)\Snn
     \Big(
     {\rho}\big(N-K_d+1\big)\sum_{m=1}^{M}\!
     {(1-a_m)\etamlE} \gamsml
     \!+
     {\rho}\!\sum_{m=1}^{M}\sum_{\ell'\neq \ell}^{L}\!
   {(1-a_m)\etamlpE} \betamls \! +\!
   {\rho}
   \sum_{m=1}^{M}\!\sum_{k=1}^{K_d}\!
   {a_m\etamkI}\betamls + 1\Big).
\end{align}
	\hrulefill
	\vspace{-5mm}
\end{figure*}
\end{Theorem}


\vspace{-1.6em}
 \section{ Problem Formulation and Proposed Solution }~\label{sec:optimization}
In this section, we formulate and solve the AP operation mode selection problem to maximize the average of sum-HE. More specifically, we aim to optimize the AP operation mode selection vectors ($\aaa$) and power control coefficients ($\ETAI, \ETAE$) to maximize the average sum-HE, subject to minimum power requirements at the EUs, per-IU SE constraints, and transmit power at each APs. The optimization problem is mathematically formulated as
\vspace{-0.1em}
\begin{subequations}\label{P:SHE:max}
	\begin{align}
		\text{\textbf{(P1):}}~\underset{\aaa, \ETAI, \ETAE}{\max}\,\, \hspace{0.5em}&
		\sum_{\ell\in \mathcal{L}} 
  \Ex\left\{\Phi_{\ell}\big(\aaa,\ETAI, \ETAE\big)\right\}
		\\
		\mathrm{s.t.} \,\,
		\hspace{0.5em}& \Ex\left\{\Phi_{\ell}\big(\aaa,\ETAI, \ETAE\big)\right\} \geq \Gamma_{\ell},~\forall \ell\in\mathcal{L},\\
		& \mathrm{SE}_k (\aaa,  \ETAI, \ETAE)  \geq \SEQoS,~\forall k \in \K_d,,\label{eq:SE:ct1}\\
			&\sum_{k=1}^{K_d}
        {\etamkI}\leq a_m,
        ~\forall m\in\MM,\label{eq:etamkI:ct1}\\
       &\sum_{\ell=1}^{L}
        {\etamlE}\leq 1-a_m,~\forall m\in\MM,\label{eq:etamellE:ct1}\\
		& a_m \in\{ 0,1\},\label{eq:am:ct1}
		\end{align}
\end{subequations}
where $\SEQoS$ is the minimum SE required by the $k$-th IU; $\Gamma_{\ell}$ is the minimum required HE at EU $\ell$.  We note that the power constraint at AP $m$ can also be written as $a_m\sum_{k=1}^{K_d}{\etamkI}+(1-a_m)\sum_{\ell=1}^{L} {\etamlE} \leq 1$. However, for instance, if $a_m=0$, this constraint can allow $\etamkI$ to take a large value which may cause some numerical issues for optimization methods to be developed. Thus, we introduce equivalent power constraints given in 
\eqref{eq:etamkI:ct1} and \eqref{eq:etamellE:ct1}. In this way, the power coefficients for information and energy transfer are forced to zero according to the AP's operating mode.

\vspace{-2em}
\subsection{Solution}
Before proceeding, by inspecting~\eqref{eq:NLEH}, we notice that $\Omega$ does not have any effect on the the optimization problem. Therefore, we directly  consider $\Psi(\mathrm{E}_{\ell}(\aaa, \ETAI, \ETAE))$ to describe the harvested energy at EU $\ell$. The inverse function of~\eqref{eq:PsiEl} can be written as
\begin{align}~\label{eq:EinvPsi}
    f (\Psi_\ell) =  \chi - \frac{1}{\zeta}\ln\Big( \frac{\phi-\Psi_{\ell}}{\Psi_{\ell}}\Big), \forall \ell. 
\end{align}

Moreover, since the logistic function in~\eqref{eq:PsiEl} is a convex function of $\mathrm{E}_{\ell}(\aaa, \ETAI, \ETAE)$, by using Jensen's inequality, we have 
\begin{align}\label{eq:Jensen}
 \Ex\left\{\Psi_{\ell}\left(\mathrm{E}_{\ell}\big(\aaa,  \ETAI, \ETAE\big)\right)\right\}&
 \geq \Psi_{\ell}\left(\Ex\left\{\mathrm{E}_{\ell}\big(\aaa, \ETAI, \ETAE\big)\right\}\right) 
 \nonumber\\
 &=\Psi_{\ell}\left(Q_{\ell}\big(\aaa,  \ETAI, \ETAE\big)\right).
 \end{align}

Now, by invoking~\eqref{eq:EinvPsi} and~\eqref{eq:Jensen}, and considering the auxiliary variables  $\BOmega\!=\!\{\omega_1,\ldots,\omega_{L}\}$, we reformulate problem~\eqref{P:SHE:max} as 
\vspace{0.2em}
\begin{subequations}\label{P:SHE:max:P2.1}
	\begin{align}
		\underset{\aaa, \ETAI, \ETAE, \BOmega}{\max}\,\, \hspace{2em}&
		\sum_{\ell\in \mathcal{L}} \omega_{\ell} 
		\\
		\mathrm{s.t.} \,\,
  		\hspace{2em}& \omega_{\ell} \geq \Gamma_{\ell},~\forall \ell\in\mathcal{L},~\label{eq:EH:ctomgell:P2}\\
           \hspace{0.5em}& Q_{\ell} (\aaa, \ETAI, \ETAE) \geq 
           f (\omega_{\ell}),~\forall \ell\in\mathcal{L},
           ~\label{eq:EH:ct:P2}\\
  &~\eqref{eq:SE:ct1}-\eqref{eq:am:ct1}.
		\end{align}
\end{subequations}
Next, we consider the continuous relaxation method to solve problem~\eqref{P:SHE:max:P2.1}. In this context, the difficulty in solving the above problem lies in the nonconvexity of \eqref{eq:EH:ct:P2} and \eqref{eq:SE:ct1}. To deal with these constraints, we apply SCA. Let us consider \eqref{eq:EH:ct:P2} first, which is equivalent to
 \vspace{-0.3em}
\begin{multline}
\rho\big(N-K_{d}+1\big)\sum_{m=1}^{M}\gamsml\etamlE+\rho\sum_{m=1}^{M}\sum_{\ell'\neq\ell}^{L}\betamls\etamlpE
\\
+\rho\sum_{m=1}^{M}a_{m}\uml+1
\geq\frac{\HEQoS}{(\tau_{c}-\tau)\Snn},
\end{multline}
where
\vspace{0.2em}
\[
\uml\triangleq\betamls\sum_{k=1}^{K_{d}}\etamkI-\big(N-K_{d}+1\big)\gamsml\etamlE-\betamls\sum_{\ell'\neq\ell}^{L}\etamlpE.
\]
Note that $\uml$ is not treated as a new optimization variable,
but as an ``expression holder''. Hence, \eqref{eq:EH:ct:P2} is now equivalent
to
\vspace{0.2em}
\begin{align}
\label{eq:ct:EH}
&4\rho\big(N-K_{d}+1\big)\sum_{m=1}^{M}\gamsml\etamlE+4\rho\sum_{m=1}^{M}\sum_{\ell'\neq\ell}^{L}\betamls\etamlpE\\
&\hspace{1em}+\!\rho\!\!\!\sum_{m=1}^{M}(a_{m}\!+\uml)^{2}\!+4\!\geq\!\frac{4\HEQoS}{\eta(\tau_{c}\!-\tau)\Snn}\!+\!\rho\!\!\sum_{m=1}^{M}\!(a_{m}\!-\uml)^{2}.
\nonumber
\end{align}
To facilitate the description we use a superscript $(n)$ to denote
the value of the involving variable produced after $(n-1)$ iterations
($n\geq0$). In light of SCA,~\eqref{eq:ct:EH} can be approximated
by the following convex one
\vspace{0.2em}
\begin{multline}
4\rho\big(N-K_{d}+1\big)\sum_{m=1}^{M}\etamlE\gamsml+4\rho\sum_{m=1}^{M}\sum_{\ell'\neq\ell}^{L}\etamlpE\betamls\\
+\rho\!\sum_{m=1}^{M}(a_{m}^{(n)}+\umln)\Bigl(2(a_{m}+\uml)-a_{m}^{(n)}-\umln\Bigr)+4\\
\geq\frac{4\tilde{f}(\omega_{\ell})}{\eta(\tau_{c}-\tau)\Snn}+\rho\!\!\sum_{m=1}^{M}(a_{m}-\uml)^{2}, \label{eq:energy:approx1}
\end{multline}
where $\umln=\betamls\sum_{k=1}^{K_{d}}{\etamkIn}-\big(N-K_{d}+1\big)\gamsml{\etamlEn}-\betamls\sum_{\ell'\neq\ell}^{L}{\etamlEn}$
and we have used the following inequality
\begin{equation}~\label{eq:x2:ineq}
x^{2}\geq x_{0}^{2}+2x_{0}(x-x_{0})=x_{0}(2x-x_{0}),
\end{equation}
and replaced $x$ and $x_{0}$ by $(a_{m}+\uml)$ and $(a_{m}^{(n)}+\uml^{(n)})$,
respectively. Moreover, we have replaced the non-convex function $\HEQoS$ with its convex upper bound, i.e,  
\begin{align}
   \tilde{f}(\omega_{\ell})
   &\!\triangleq \!\chi \!-\frac{1}{\zeta}\Big(\!
   \ln\big( \phi\!-\omega_{\ell}\big) \!- \!\Big( \ln\big(\omega_{\ell}^{(n)}\big) \!+ \!\frac{1}{\omega_{\ell}^{(n)}}(\omega_{\ell}\!-\omega_{\ell}^{(n)}) \Big)\Big).\nonumber
\end{align}

Now, we turn our attention to \eqref{eq:SE:ct1} which is equivalent to
\vspace{-0.5em}
\begin{multline}~\label{eq:SE:ct:expnd}
\frac{\rho\big(N-K_{d}\big)}{2^{\barSEQoS}-1}
\Big(\sum_{m=1}^{M}\sqrt{a_{m}\etamkI\gamuemk}\Big)^{2}\geq\\
\sum_{m=1}^{M}\rho\nu_{mk}a_{m}\left(\eta_{m}^{\mathtt{I}}-\eta_{m}^{\mathtt{E}}\right)+\sum_{m=1}^{M}\sum_{\ell=1}^{L}\rho\nu_{mk}\etamlE+1,
\end{multline}
where $\barSEQoS=\frac{\SEQoS}{1- {\tau}/{\tau_c}}$, $\nu_{mk}=\betamkue\!-\!\gamuemk$, $\eta_{m}^{\mathtt{I}}\triangleq\sum_{k'=1}^{K_{d}}\etamkpI$,
and $\eta_{m}^{\mathtt{E}}\triangleq\sum_{\ell=1}^{L}\etamlE$. 
We can further easily rewrite~\eqref{eq:SE:ct:expnd} as 
\vspace{-0.2em}
\begin{multline}~\label{eq:SE:ct:expnd2}
\hspace{-0.2em}\frac{\rho\big(N\!-\!K_{d}\big)}{2^{\barSEQoS}-1}
\!\Big(\!\!\sum_{m=1}^{M}\!\!\sqrt{a_{m}\etamkI\gamuemk}\Big)^{\!2}\!+\frac{1}{4}\!\!\sum_{m=1}^{M}\!\!\rho\nu_{mk}\bigl(a_{m}+\eta_{m}^{\mathtt{I}}-\eta_{m}^{\mathtt{E}}\bigr)^{\!2}\\
\geq\frac{1}{4}\sum_{m=1}^{M}\rho\nu_{mk}\bigl(a_{m}-\eta_{m}^{\mathtt{I}}+\eta_{m}^{\mathtt{E}}\bigr)^{2}+\sum_{m=1}^{M}\rho\nu_{mk}\eta_{m}^{\mathtt{E}}+1.
\end{multline}
Now, we need to find a concave lower bound of the left
hand side of~\eqref{eq:SE:ct:expnd2}. To this end, by invoking~\eqref{eq:x2:ineq}, we have
\vspace{-0.2em}
\begin{align}
\Big(\sum_{m=1}^{M}\sqrt{a_{m}\etamkI\gamuemk}\Big)^{2}&\geq q_{k}^{(n)}\Bigl(2\sum_{m=1}^{M}\sqrt{\gamuemk a_{m}\etamkI}-q_{k}^{(n)}\Bigr),\nonumber\\
z_{m}^2 &\geq z_{m}^{(n)}\bigl(2z_{m}-z_{m}^{(n)}\bigr),
\end{align}
where $q_{k}^{(n)}=\sum_{m=1}^{M}\sqrt{\gamuemk a_{m}^{(n)}\etamkIn}$, $z_{m}=a_{m}+\eta_{m}^{\mathtt{I}}-\eta_{m}^{\mathtt{E}}$, and $z_{m}^{(n)}=a_{m}^{(n)}+\etamIn-\etamEn$.
Therefore, \eqref{eq:SE:ct:expnd2} can be approximated by the following constraint
\vspace{-.4em}
\begin{multline}
\frac{\rho\big(N-K_{d}\big)}{2^{\barSEQoS}-1}q_{k}^{(n)}\biggl(2\sum_{m=1}^{M}\sqrt{\gamuemk a_{m}\etamkI}-q_{k}^{(n)}\biggr)\\
+\frac{1}{4}\sum_{m=1}^{M}\rho\nu_{mk}z_{m}^{(n)}\bigl(2z_{m}-z_{m}^{(n)}\bigr)\\
\geq\frac{1}{4}\sum_{m=1}^{M}\rho\nu_{mk}\bigl(a_{m}-\eta_{m}^{\mathtt{I}}+\eta_{m}^{\mathtt{E}}\bigr)^{2}+\sum_{m=1}^{M}\rho\nu_{mk}\eta_{m}^{\mathtt{E}}+1.\label{eq:SINR:approx:SHE1}
\end{multline}

To this end, we now arrive at the following approximate convex problem
\vspace{-.5em}
\begin{subequations}\label{P:SHE:max:final}
	\begin{align}
		\underset{\aaa, \ETAI, \ETAE,\BOmega}{\max}\,\, \hspace{2em}&
		\sum_{\ell\in \mathcal{L}} \omega_{\ell} 
		\\
		\mathrm{s.t.} \,\,
  		\hspace{2em} 
         & ~\eqref{eq:energy:approx1},\eqref{eq:SINR:approx:SHE1}~~\forall \ell\in\mathcal{L}, \forall k \in \K_d, 
                \\                &~\eqref{eq:EH:ctomgell:P2},\eqref{eq:etamkI:ct1},\eqref{eq:etamellE:ct1}.
		\end{align}
\end{subequations}
The convex optimization problem~\eqref{P:SHE:max:final} can be efficiently solved by using existing standard convex optimization packages, such as CVX~\cite{cvx}. We iteratively solve problem~\eqref{P:SHE:max:final} until the relative reduction of the objective function's value in~\eqref{P:SHE:max:final} falls below the predefined threshold, $\epsilon$. 

\emph{Complexity Analysis:}  In each iteration of solving the optimization problem~\eqref{P:SHE:max:final}, the computational complexity is $\mathcal{O} (\sqrt{A_l + A_q} (A_l + A_q + A_v)A_v^2)$, where $A_l=2M+L$ denotes the number of linear constraints, $A_q = K_d + L$ presents the number of quadratic constraints and $A_v = M(K_d+L+1) +L$ stands for the number of real-valued scalar variables~\cite{tam16TWC}.

\vspace{-1em}
\subsection{Benchmarks}
\label{sec:benchmark}
To investigate the effectiveness  of the proposed scheme 
we compare it against the following benchmarks.
\subsubsection{Random AP Operation Mode Selection without Power Control (Benchmark 1)}
We assume that the APs' operation mode selection parameters ($\aaa$) are randomly assigned and no power control is performed at the APs. PZF and PMRT precoding  are applied to the I-APs and E-APs, respectively. Moreover, in the absence of power control, both E-APs and I- APs transmit at full power, i.e., at the $m$-th AP, power coefficients are the same and $\etamkI = \frac{1}{K_d}$, $\forall k\in\mathcal{K}_d$  and $\etamlE = \frac{1}{L}$, $\forall \ell\in\mathcal{L}$. 
\subsubsection{Random AP Operation Mode Selection with Power Control (Benchmark 2)}
Assuming random AP operation mode selection, we optimize the power control coefficients ($\ETAI$, $\ETAE$), under the same SE requirement constraints for IU and energy requirements for EUs as in the proposed scheme. It is important to  note that, if $a_m=0$ we need to set $\etamkI = 0$ for the associated APs. More specifically, we remove  $\etamkI $ from the optimization and only consider $\etamlE$ for these APs. Similarly, if $a_m=1$ we should have $\etamlE = 0$ and these APs only have $\etamkI$. Based on these observations, the optimization problem \textbf{(P2)} is reduced to
\vspace{-0.8em}
\begin{subequations}\label{P:SHE:max:S1}
	\begin{align}
		\text{\textbf{(P2-1):}}~\underset{\ETAI, \ETAE}{\max}\,\, \hspace{0.5em}&
		\sum_{\ell\in \mathcal{L}}  \Ex \left\{\Phi_{\ell}^{\Son} ( \ETAI, \ETAE)\right\}
		\\
		\mathrm{s.t.} \,\,
		\hspace{0.5em}&  \Ex\left\{\Phi_{\ell}^{\Son} (\ETAI, \ETAE) \right\}\geq \Gamma_{\ell},~\forall \ell\in\mathcal{L},\\
		& \mathrm{SE}_k^{\Son} ( \ETAI, \ETAE)  \geq \SEQoS,~\forall k \in \K_d,\\
			&\sum_{k=1}^{K_d}
        {\etamkI}\leq 1,
        ~\forall m\in\MM \textrm{with } a_m=1,\label{eq:infopower}\\
       &\sum_{\ell=1}^{L}
        {\etamlE}\leq 1,~\forall m\in\MM \textrm{with } a_m=0,\label{eq:powerenergy}
		\end{align}
\end{subequations}
where $\SEk^{\Son}( \ETAI, \ETAE)$ and $\Phi_{\ell}^{\Son} ( \ETAI, \ETAE)$ are given in~\eqref{eq:SEk:Ex} and~\eqref{eq:NLEH}, respectively, for a given $\qa$. Problem~\eqref{P:SHE:max:S1} has the same structure as problem~\eqref{P:SHE:max}. Therefore, we use the same solution with some slight modifications.

\subsubsection{SWIPT with Orthogonal Transmission (Benchmark 3)} 
We assume that all APs are used for DL WIT and WPT over orthogonal time frames. More specifically, the duration of $(\tau_c-\tau)$ is divided into two equal time fractions of length $(\tau_c-\tau)/2$. In the first fraction, all APs transmit DL information for the IUs by using the PZF precoding and in the consecutive fraction, all APs transmit energy symbols towards the EUs via MRT precoding. Thus, the ergodic SE for the $k$-th IU  and average HE for the $\ell$-th EU are respectively given by 
\vspace{0em}
\begin{align}
    \mathrm{SE}_k^{\mathtt{S}_2}(\ETAI)
     & =
      \frac{1}{2}\Big(1\!- \!\frac{\tau}{\tau_c}\Big)
      \nonumber\\
      &\hspace{-4em}
      \times\log_2
      \Bigg(
       1\! + \!\frac{
                  \rho \big(N-K_d\big)\Big(\sum_{m=1}^{M}\sqrt{ \etamkI \gamuemk}  \Big)^2
                 }
                 { \sum_{m=1}^{M}
                 \sum_{k'=1}^{K_d}
  \rho \etamkpI 
  \Big(\betamkue-\gamuemk\Big)
                   \!+\!  1}
     \Bigg),~\label{eq:SEk:S2}\\
    Q_{\ell}^{\mathtt{S}_2} (\ETAE)
 &=
     \frac{(\tau_c-\tau)}{2} \Snn
     \Big(
     {\rho}\big(N+1\big)\sum_{m=1}^{M}
     {\etamlE} \gamsml
     +
     \nonumber\\
     &\hspace{2em}
     {\rho}\!\sum_{m=1}^{M}\sum_{\ell'\neq \ell}^{L}\!
   {\etamlpE} \betamls +1\Big).~\label{eq:El:S2}
\end{align}

Since $\mathrm{SE}_k^{\mathtt{S}_2}(\ETAI)$ only depends on $\ETAI$ and $ Q_{\ell}^{\mathtt{S}_2} (\ETAE)$ is only a function of $\ETAE$, by applying SCA, the power control problem can be decoupled into two separate problems
\begin{subequations}\label{P:SHE:max:S2:R1}
	\begin{align}
		\text{\textbf{(P2-2a):}}~\underset{\ETAE, \BOmega^{\Sto}}{\max}\,\, \hspace{0.5em}&
		\sum_{\ell\in \mathcal{L}} 
 \omega_{\ell}^{\Sto}
		\\
		\mathrm{s.t.} \,\,
		\hspace{0.5em}& \omega_{\ell}^{\Sto}\geq \Gamma_{\ell},~\forall \ell\in\mathcal{L},\\
           \hspace{2em}& 
           {\rho}\big(N+\!1\big)\!\sum_{m=1}^{M}\!
     {\etamlE} \gamsml
     \!+\!
     {\rho}\!\sum_{m=1}^{M}\sum_{\ell'\neq \ell}^{L}\!
   {\etamlpE} \betamls \!+\! 1 \nonumber\\
   &\geq 
    \frac{2\tilde{f}(\omega_{\ell}^{\Sto})}  {(\tau_c-\tau)\Snn}, ~\forall \ell\in\mathcal{L}           
                     ~\label{eq:EH:ct::S2:R1}\\
       &\sum_{\ell=1}^{L}
        {\etamlE}\leq 1,~\forall m\in\MM,\label{eq:powerenergy:S2:R1a}
		\end{align}
\end{subequations}
\begin{subequations}\label{P:SHE:max:S2:R1}
	\begin{align}
		\text{\textbf{(P2-2a):}}~{\text{Find}}\,\, \hspace{0.5em}&
		\ETAI
		\\
		\mathrm{s.t.} \,\,
		\hspace{0.5em}&  
       \!\frac{\rho \big(N-K_d\big)}{ 2^{\tilSEQoS}-1}
                  q_{k,\Sto}^{(n)}\bigg(2\sum_{m=1}^{M}\sqrt{ \etamkI \gamuemk}- q_{k,\Sto}^{(n)} \bigg)
                  \nonumber\\
                  &\hspace{-5em}
    \geq  \sum_{m=1}^{M}
                 \sum_{k'=1}^{K_d}
  \rho \etamkpI 
  \Big(\betamkue-\gamuemk\Big)
                   \!+\!  1,~\forall k \in \K_d,\\
			&\sum_{k=1}^{K_d}
        {\etamkI}\leq 1,
        ~\forall m\in\MM,\label{eq:infopower:S2:R1b}
		\end{align}
\end{subequations}
where $\BOmega^{\Sto} =\{\omega_1^{\Sto},\ldots,\omega_{L}^{\Sto}\}$ are auxiliary variables $\tilSEQoS=\frac{2\SEQoS}{1- {\tau}/{\tau_c}}$ and $q_{k,\Sto}^{(n)}=\sum_{m=1}^{M}\sqrt{ \etamkIn \gamuemk} $. These two sub-problems can be iteratively solved using CVX~\cite{cvx}.
 

\begin{figure}[t]
	\centering
	\vspace{-0.25em}
	\includegraphics[width=0.40\textwidth]{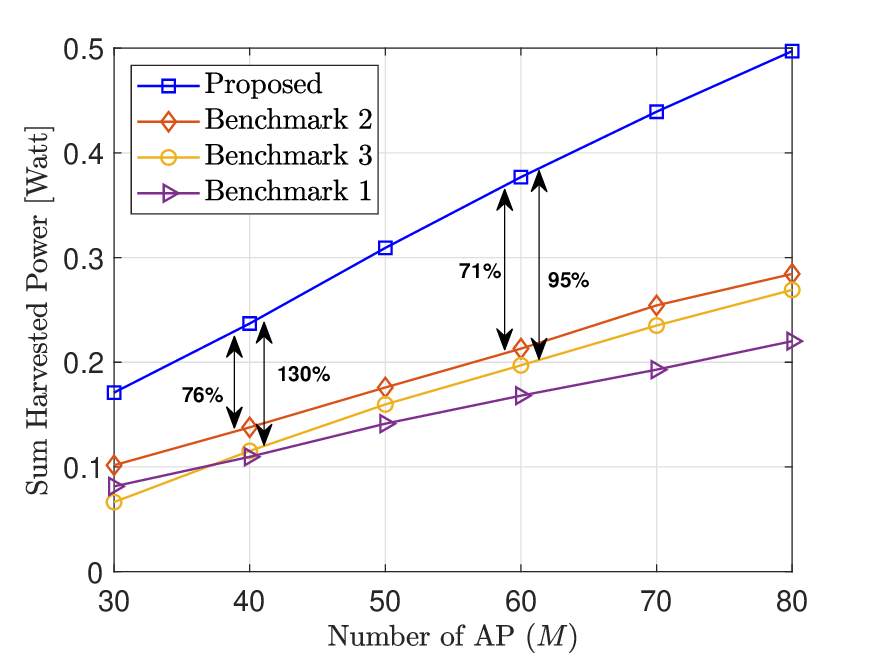}
	\vspace{-0.7em}
	\caption{Impact of the number of APs on the average of sum harvested power ($N=10$, $K_d=3$, $L=5$, $\Gamma_{\ell}=100$ $\mu$W,  and $\SEQoS = 1$ bit/s/Hz).}
	\vspace{0.4em}
	\label{fig:Fig2}
\end{figure}
\vspace{-0.8em}
\section{Numerical Examples}~\label{Sec:Numer}
Numerical results are now provided to illustrate the gains of our proposed optimization framework.  We consider a CF-mMIMO network, where the APs and UEs are randomly distributed in a square of $0.5 \times 0.5$ km${}^2$, whose edges are wrapped around to avoid the boundary effects.  We choose $\mathcal{S}_\dl^o=1$ bit/s/Hz, $\tau_c=200$,  $\tau=K_d+L$ and $\epsilon=10^{-5}$.  We set the noise power $\Sn=-92$ dBm. Let $\tilde{\rho} = 1$ W and $\tilde{\rho}_t = 0.2$~W be the maximum transmit power of the APs and UL training pilot sequences, respectively. The normalized maximum transmit powers ${\rho}$ and ${\rho}_t$ are calculated by dividing these powers by the noise power. The non-linear energy harvesting parameters are set as $\xi=150$, $\chi=0.014$, and $\phi=0.024$ Watt~\cite{Boshkovska:CLET:2015}. We model the large-scale fading coefficients as $\beta_{mk} = 10^{{\text{PL}_{mk}^d}/{10}}10^{{F_{mk}}/{10}}$  ($\beta_{mk}\in\{\betamls,\betamkue\}$) where $10^{{\text{PL}_{mk}^d}/{10}}$ represents the path loss, and $10^{{F_{mk}}/{10}}$ represents the shadowing effect with $F_{mk}\in\mathcal{N}(0,4^2)$ (in dB)~\cite{emil20TWC}.  Here, $\text{PL}_{mk}^d$ (in dB) is given by $\text{PL}_{mk}^d = -30.5-36.7\log_{10}\big({d_{mk}}/{1\,\text{m}}\big)$ and the correlation among the shadowing terms from the AP $m, \forall m\in\mathcal{M}$ to different UEs $k\in\mathcal{K}_d$ ($\ell\in\mathcal{L}$) is expressed as~\cite[Eq. (40)]{emil20TWC}.

Figure~\ref{fig:Fig2} shows the average sum harvested power achieved by the proposed scheme and the benchmark schemes as a function of the number of APs. We observe that our proposed scheme yields substantial performance gains in terms of energy harvesting efficiency over the Benchmarks, especially when the number of APs is small. This highlights the importance of joint AP operation mode selection and power control design in the proposed architecture, as Benchmarks 1 and 3 result in almost the same performance.

\begin{figure}[t]
	\centering
	\vspace{-0.25em}
	\includegraphics[width=0.40\textwidth]{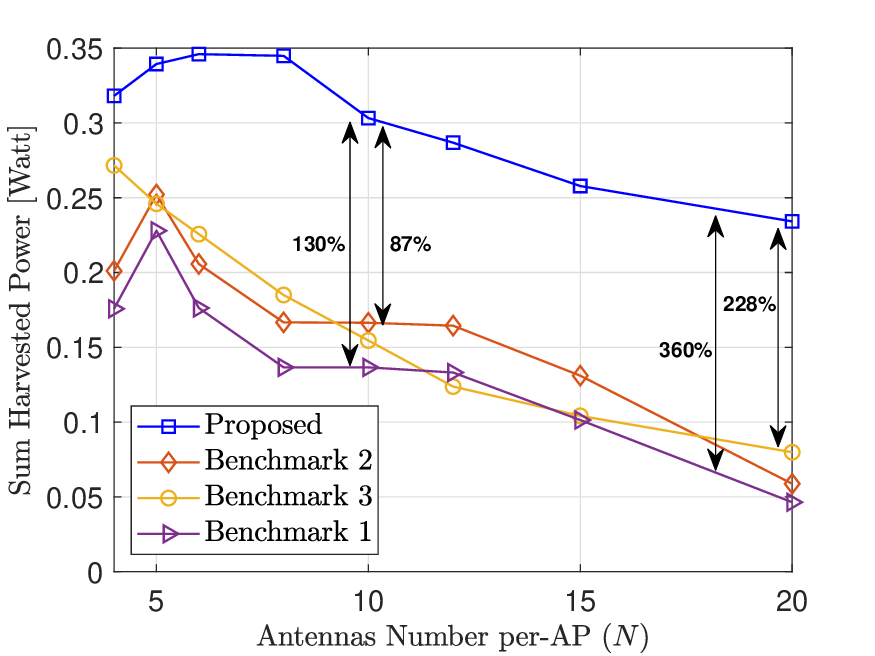}
	\vspace{-0.7em}
	\caption{Impact of the number of AP antennas on the average of sum harvested power ($NM=480$, $K_d=3$, $L=5$, $\Gamma_{\ell}=100$ $\mu$W, and $\SEQoS = 1$ bit/s/Hz).}
	\vspace{0.4em}
	\label{fig:Fig3}
\end{figure}

Figure~\ref{fig:Fig3} illustrates the average sum harvested power versus the number of antennas per AP. We note that, for a fixed number of service antennas, the number of APs decreases as the number of antennas per AP $N$ increases. The increase in the number of antennas per AP provides more degrees-of-freedom for energy harvesting at the EUs. However, the distance between the E-APs and EUs increases as a consequence of  the decreased number of APs,  and thus, the detrimental effects due to severe path loss diminish  the benefits provided by the increased number of antennas per AP. Therefore, the sum-HE first increases, approaches the optimal point, and then decreases as $N$ increases. This trend is also  observed for Benchmarks 1 and 2, while the average sum-HE by Benchmark 3 is a monotonically decreasing function of $N$. Finally, when both the number of APs and the number of antennas per-AP become large, Benchmark 2 outperforms Benchmark 3, indicating that the former can be used for large $N$  when computational complexity is a concern.  

\vspace{-0.5em}
\section{Conclusion}~\label{Sec:conc}
We investigated the problem of sum-HE maximization in CF-mMIMO systems with separate IUs and EUs. We proposed a novel architecture, where, based on the network requirements, the AP operation mode and the associated  power control coefficients were jointly  optimized.  Numerical results revealed that the proposed architecture offers significant boost to energy harvesting efficiency, especially with a smaller number of APs and a larger number of AP antennas. Moreover, with a fixed number of service antennas, there is an optimum setup for the number of APs and per-AP antenna, yielding maximum HE.  

\vspace{0em}
\bibliographystyle{IEEEtran}


\end{document}